\documentstyle[preprint,prd,aps]{revtex}
\begin{document}

%
 
\let\a=\alpha      \let\b=\beta       \let\c=\chi        \let\d=\delta   
\let\e=\varepsilon \let\f=\varphi     \let\g=\gamma      \let\h=\eta       
\let\k=\kappa      \let\l=\lambda     \let\m=\mu               
\let\o=\omega      \let\r=\varrho     \let\s=\sigma    
\let\t=\tau        \let\th=\vartheta  \let\y=\upsilon    \let\x=\xi       
\let\z=\zeta       \let\io=\iota      \let\vp=\varpi     \let\ro=\rho        
\let\ph=\phi       \let\ep=\epsilon   \let\te=\theta  
\let\n=\nu  
\let\D=\Delta   \let\F=\Phi    \let\G=\Gamma  \let\L=\Lambda  
\let\O=\Omega   \let\P=\Pi     \let\Ps=\Psi   \let\Si=\Sigma  
\let\Th=\Theta  \let\X=\Xi     \let\Y=\Upsilon  
 
%
 
%
 
\def\cA{{\cal A}}                \def\cB{{\cal B}} 
\def\cC{{\cal C}}                \def\cD{{\cal D}} 
\def\cE{{\cal E}}                \def\cF{{\cal F}} 
\def\cG{{\cal G}}                \def\cH{{\cal H}}                  
\def\cI{{\cal I}}                \def\cJ{{\cal J}}                 
\def\cK{{\cal K}}                \def\cL{{\cal L}}                  
\def\cM{{\cal M}}                \def\cN{{\cal N}}                 
\def\cO{{\cal O}}                \def\cP{{\cal P}}                 
\def\cQ{{\cal Q}}                \def\cR{{\cal R}}                 
\def\cS{{\cal S}}                \def\cT{{\cal T}}                 
\def\cU{{\cal U}}                \def\cV{{\cal V}}                 
\def\cW{{\cal W}}                \def\cX{{\cal X}}                  
\def\cY{{\cal Y}}                \def\cZ{{\cal Z}} 
%
 
\newcommand{\Ns}{N\hspace{-4.7mm}\not\hspace{2.7mm}} 
\newcommand{\qs}{q\hspace{-3.7mm}\not\hspace{3.4mm}} 
\newcommand{\ps}{p\hspace{-3.3mm}\not\hspace{1.2mm}} 
\newcommand{\ks}{k\hspace{-3.3mm}\not\hspace{1.2mm}} 
\newcommand{\des}{\partial\hspace{-4.mm}\not\hspace{2.5mm}} 
\newcommand{\desco}{D\hspace{-4mm}\not\hspace{2mm}}


\draft
\title{Rare K decays in a model of quark and lepton masses} 
\author{P. Q. Hung\cite{email1}, Andrea Soddu\cite{email2}}
\address{Dept. of Physics, University of Virginia, \\
382 McCormick Road, P. O. Box 400714, Charlottesville, Virginia 22904-4714}
\date{\today}
\maketitle

\begin{abstract}
An extension of a model of neutrino masses to the quark sector provides
an interesting link between these two sectors. A parameter which is important
to describe neutrino oscillations and masses is found to be a crucial one
appearing in various ``penguin'' operators, in particular 
the so-called Z penguin. This parameter is severely constrained by the
rare decay process $K_{L} \rightarrow \mu^{+} \mu^{-}$. This in turn has
interesting implications on the decay rates of other rare processes such as
$K_{L} \rightarrow \mu e$, etc..., as well as on the masses of the neutrinos
and the masses of the vector-like quarks and leptons which appear in
our model.
\end{abstract}

\pacs{12.15.Ff, 12.60.Cn, 13.20.Eb, 14.60.Pq, 14.60.St, 14.70.Pw}

\section{Introduction}
In the last few years, we have witnessed a flurry of far-reaching
experimental results, among which are the neutrino oscillation 
data \cite{superk},  data on direct CP-violation such as $\epsilon^{\prime}
/\epsilon$ \cite{epsilon} 
and upper bounds on Flavour-Changing-Neutral-Current (FCNC) 
rare decays of the kaons. On the one hand, 
the neutrino oscillation data clearly
points to possible physics beyond the Standard Model (SM). On the
other hand, it is still not clear if the new results on
$\epsilon^{\prime}/\epsilon$, which differ roughly by a factor of two
from present calculations within the SM, imply any new physics
since the aforementioned calculations are still plagued with
non-perturbative uncertainties. For the kaon's FCNC rare decays, the 
experimental situation is still far from  giving evidences of physics beyond 
the SM or to confirm the SM itself. Nevertheless, whatever new physics,
which might be responsible for giving rise to neutrino masses,
could, in principle, affect the quark sector, and hence, also
quantities such as $\epsilon^{\prime}/\epsilon$ or the branching ratios of the 
kaon's FCNC rare decays . If this is the case,
results from the quark sector could then be used to put constraints
on the lepton sector itself since it is possible that both sectors
have a common set of parameters, a desirable feature of any model
which purports to deal with issues of fermion masses. The aim of this
paper is to explore this possible connection between the two
sectors.

The subject of neutrino masses has been invigorated in the last
few years due to new results on neutrino oscillations which
suggested the possibility of a non-vanishing mass for the neutrinos.
Models have been (and are still being) built to try to describe
these oscillations. In a large number of cases, efforts
were mainly concentrated on the type of neutrino mass matrices which
could ``explain'' the oscillation data, with very little attempt
made in trying to connect that kind of physics to the hadronic
sector. However, it is perfectly reasonable to expect that the
two sectors are somehow deeply connected, and that a constraint from
one sector can give rise to constraints on the other sector.
This particular connection will be the subject of the present
manuscript. This paper is an extension to the quark sector
of a model of neutrino mass presented in Ref. \cite{PQ}. In
consequence, we shall show that there is a set of parameters
which appear in both sectors, and that the constraints obtained
in the quark sector have interesting implications
on the neutrino sector. In particular, we shall use recent
results from limits on various rare decays to constrain a common parameter
which appears in both sectors.
( As we will show, because this common parameter is real, 
there will be no new 
contributions from our model to $\epsilon^{\prime}/\epsilon$.)

The plan of this paper is as follows. First, we present our model
for the quark sector based on a previous model for the neutrino
sector \cite{PQ}. We then proceed to enumerate and compute
various FCNC operators which
arise in this model. These operators are important in the
analysis of various rare decays. 
Finally, we will use these FCNC operators in the
computation of their contributions to the aforementioned
quantities, and set constraints on the parameters of the model.
Using these constraints, we look at the question of how they
would affect the neutrino sector. We will also show at the end
of the paper that there are some interesting correlations
between the value of the branching ratio for
$K_{L} \rightarrow \mu \,e$ and the mass of the
weak-singlet quarks and charged leptons which appear in
our model. When these particles are ``light'' enough to be
produced at future accelerators, the branching ratio for
$K_{L} \rightarrow \mu \,e$ is too small, of 
order $10^{-22}$, to be detected, while for a branching
ratio which could conceivably be measured in a not-too-distant
future, e.g. of order $10^{-14}$, the masses of these singlet
quarks and leptons will be too large, a few hundreds of TeVs
or more, to be produced by earthbound laboratories.

\section{ Effective vertices for FCNC processes in the model of \\
\protect \hspace{-12.2cm} Ref.\protect \cite{PQ}  }

The model used in Ref. \cite{PQ} to describe the lepton sector is 
summarized in Appendix A. (A quick look at that appendix will help
with the notations and particle content.)
In this section, we will simply write down the 
part of the Lagrangian for the quark sector which is
relevant to the construction of various FCNC operators. First of all,
the common link between the quark and lepton sectors is in the scalar
sector. These are the SM Higgs field $\phi$ and the Family symmetry
Higgs field $\Omega^{\alpha}$. As we shall see below, it is these
scalars which ``transmit information'' from the quark to the
lepton sector and vice versa. In particular, as can be seen from
Appendix A, $\Omega^{\alpha}$ provides the 
common set of parameters which appears in the two sectors.
The generalization from the lepton sector to the quark sector
necessitates the introduction of a new set of vector-like
quarks, $F^{i}$, ${\cal M}_1^{i}$, and ${\cal M}_2^{i}$,
where $i$ is a color index,
in perfect analogy with the leptonic vector-like fermions,
$F$, ${\cal M}_1$, and ${\cal M}_2$. As we shall see below,
these vector-like quarks can have masses as low as a couple
of hundred GeV's, making them very attractive for potential
discoveries at the LHC. We shall come back to this interesting issue
in the last section of the manuscript.

The particle content and quantum numbers of the model are listed in Table 1. 
Notice that, among the scalars listed in that table, there is one
which does not have the appropriate quantum numbers to
be able to couple directly to the quarks: $\rho^{\alpha}$. The
gauge structure of the model (for both quarks and leptons) is
given by:
\begin{equation}
SU(3)_c \otimes SU(2)_L \otimes U(1)_Y \otimes SO(4) \otimes SU(2)_{\nu_R} \, ,
\label{gaugegroup}
\end{equation}

\noindent
where $SU(2)_{\nu_R}$ applies only to the
right-handed neutrinos, and $SO(4)$ is the family symmetry.
The form of the quark Yukawa Lagrangian 
is similar to the one from the lepton sector
(Appendix A), with the introduction of the following new parameters:

\begin{equation}
 g_u,g_d,G_1,G_2,G_3,G_{{\cal M}_1},G_{{\cal M}_2},M_F,M_{{\cal M}_1},
M_{{\cal M}_2}  \, .
\end{equation}

\noindent
The quark Yukawa Lagrangian takes the form

\begin{eqnarray}     
{\cal L}_{quark}^{Y} & = & 
             g_{d}\bar{q}_{L}^{\a}{\phi}d_{\a R} + 
             g_{u}\bar{q}_{L}^{\a}{\tilde{\phi}}u_{\a R} + \nonumber  \\
                     &   &  
             G_{1}\bar{q}_{L}^{\a}{\Omega}_{\a}F_{R} + 
             G_{{\cal M}_1}\bar{F}_{L}{\phi}{\cal M}_{1 R} +  
             G_{{\cal M}_2}\bar{F}_{L}{\tilde{\phi}}{\cal M}_{2 R} +   
\nonumber   \\
                     &   &  
             G_{2}\bar{{\cal M}}_{1 L}{\Omega}_{\a}d_{R}^{\a} + 
             G_{3}\bar{{\cal M}}_{2 L}{\Omega}_{\a}u^{\a}_{R} +  \nonumber \\
                     &   &  
             M_{F}\bar{F}_{L}F_{R} + 
             M_{{\cal M}_1}\bar{{\cal M}}_{1 L}{\cal M}_{1 R} + 
                                                            \nonumber      \\
                     &   &  
            M_{{\cal M}_2}\bar{{\cal M}}_{2 L}{\cal M}_{2 R}  + h.c. \, , 
\end{eqnarray} 

\noindent
with the important difference with respect to the lepton sector in that 
there is no coupling between the quarks and the scalar $\rho^{\alpha}$ before
symmetry breaking. After symmetry breaking, the mass eigenstates are linear
combinations of $\Omega$ and $\rho$, and vice versa, as shown below and in 
Appendix A.

The way the FCNC processes take place in the quark sector is identical to 
the way they operate in the lepton sector i.e. via loop diagrams. 
In the SM, as it is well known, the flavour diagonal structure 
of the 
basic vertices involving $\gamma$, $Z$ and $g$ forbids the appearance of FCNC 
processes at the tree level. 
However the FCNC processes can happen at one loop or higher order level, 
mediated by a combination of flavour changing charged currents coupled to 
the W's.
The fact that these processes take place only as loop effects makes them 
particularly useful for testing the quantum structure of the theory and in 
search for physics beyond the SM.
In our model, beside the ways in which FCNC processes can happen in the SM, 
it can also happen because of the couplings of quarks of different 
flavour to the same vector-like fermion F and to the NG bosons 
$\tilde{\Omega}_i$ and the pseudo NG bosons $Re \tilde{\rho}_i$.
This is made possible by the mixings among the NG bosons $\tilde{\Omega}_i$ 
and the pseudo NG 
bosons $Re \tilde{\rho}_j$ with different family indices $i$ and $j$. 
We denote the relevant scalar mass eigenstates by $\tilde{\Omega}_i$ 
and $Re \tilde{\rho}_i$ , in terms of which we can express the states 
entering the Yukawa Lagrangian by

\begin{eqnarray}
\Omega_i & = & \cos{\b} \tilde{\Omega}_i - \sin{\b}Re \tilde{\rho}_i \, ,\\
Re \rho_i   & = & \sin{\b} \tilde{\Omega}_i + \cos{\b}Re \tilde{\rho}_i \, . 
\end{eqnarray}

\noindent
The states $\tilde{\Omega}_i$ are the NG bosons which are absorbed by the 
corresponding family gauge bosons. 
When the NG bosons get mixed, there will be mass mixings among the 
corresponding family gauge bosons.
If we denote  by $A_{\Omega}$ the orthogonal matrix which diagonalizes the
family gauge bosons mass matrix (Ref. \cite{PQ}), we can express 
the states $\tilde{\Omega}_i^{\prime}$, corresponding to the longitudinal 
components of the gauge boson mass 
eigenstates, in terms of the mass eigenstates $\tilde{\Omega}_i$

\begin{equation}
\tilde{\Omega}_i = A^T_{\Omega,ij}\tilde{\Omega}_j^{\prime} \, ,
\end{equation}

\noindent 
with $A_{\Omega}$ given by

\begin{equation}
A_{\Omega} = \left(\begin{array}{cccc}
\frac{1}{\sqrt{2}}  & \frac{1}{\sqrt{2}} & 0 \\
-\frac{1}{\sqrt{2}} & \frac{1}{\sqrt{2}} & 0 \\
 0                 &  0                & 1 
\end{array}\right) \, .
\label{matrixA} 
\end{equation}

This mechanism, introduced to describe neutrino mass splitting in the neutral 
lepton sector, gives the following 
terms in the Yukawa Lagrangian:

\begin{eqnarray}
G_{1}\bar{q}_{L}^{i}{\Omega}_{i}F_{R} 
& = & 
\cos{\b}G_{1}\bar{q}_{L}^{i}\tilde{\Omega}_{i}F_{R} - 
\sin{\b}G_{1}\bar{q}_{L}^{i}Re\tilde{\rho}_{i}F_{R} \nonumber \\
& = &               
\cos{\b} G_{1}\bar{q}_{L}^{i}A^{T}_{\Omega,ij}
\tilde{\Omega}_{j}^{\prime}F_{R} -
\sin{\b}G_{1}\bar{q}_{L}^{i}A^{T}_{\Omega,ij}Re\tilde{\rho}_j^{\prime}F_{R} 
\, ,
\end{eqnarray}

\begin{eqnarray}
G_{2}\bar{{\cal M}}_{1 L}{\Omega}_{i}d^{i}_{R}  
& = & 
\cos{\b}G_{2}\bar{{\cal M}}_{1 L}\tilde{\Omega}_id^{i}_{R} - 
\sin{\b}G_{2}\bar{{\cal M}}_{1 L}Re\tilde{\rho}_id^{i}_{R}   \nonumber \\
& = &
\cos{\b}G_{2}\bar{{\cal M}}_{1 L}
A^{T}_{\Omega,ij}{\Omega}_{j}^{\prime}d_{R}^{i} - \sin{\b}G_{2}
\bar{{\cal M}}_{1 L}A^{T}_{\Omega,ij}Re\tilde{\rho}_j^{\prime}d_{R}^{i} 
\, ,
\end{eqnarray}

\noindent 
where we assume that the same $A_{\Omega}$ diagonalizes also the 
mass matrix of the pseudo NG  $Re\tilde{\rho}_i$ (Ref. \cite{PQ}).
It is important to notice here that $H_4$ and 
$h_4$, whose mass eigenstates are given by 
$\tilde{H}_4$ and $\tilde{h}_4$, see Eq. \ref{H4h4}, are not coupled to the 
quarks of the first three families. This implies that they will not propagate 
in the loops of the diagrams describing processes with external quarks of 
the first three generations. In this paper we will look only at processes 
involving the quarks of the first three families and so we will not care about 
the presence in our model of $H_4$ and $h_4$, where  

\begin{eqnarray}
\tilde{H}_4 & = & \cos \alpha\, H_4 + \sin \alpha\, h_4 \nonumber \, , \\
\tilde{h}_4 & = & -\sin \alpha\, H_4 + \cos \alpha\, h_4 \, .
\label{H4h4}
\end{eqnarray}

In the following we will present the expressions of the effective vertices 
for FCNC processes mediated by $ Z $,  $\gamma$, and   $g$. 
There is no extra effective vertex with 
the SM Higgs in our model because the vector-like fermions
do not couple to the SM Higgs field. We will also compare these expressions 
with the corresponding ones from the 
SM. 

For all the effective vertices, the expressions  will be given in terms of 
the linear combination 

\begin{equation}
{\cal M} = {\cal M}_{\Omega} \cos^2{\b} + {\cal M}_{\rho} \sin^2{\b} \, ,
\end{equation}

\noindent
where ${\cal M}_{\Omega}$ and ${\cal M}_{\rho}$ are the contributions to the 
effective vertices when the particles propagating in the loops are 
respectively 
the NG bosons $\tilde{\Omega}_i^{\prime}$ and the pseudo NG bosons 
$Re\tilde{\rho}_i^{\prime}$.  
From the lepton sector we have estimated $\tan{\b}$ to be 

\begin{equation}
\tan{\b} \equiv \frac{V^{\prime}}{V} \approx g_{F}^2
\frac{M_F^L M_{{\cal M}_2}^L}{M_G^2}
\, , 
\end{equation}

\noindent
with $g_{F}$ being the $SO(4)$ gauge coupling. $V^{\prime}$ and $V$ 
are the vacuum expectation values (VEV) of $\rho$ 
and $\Omega$ respectively, $M_F^L$ and $M_{{\cal M}_2}^L$ the
masses of the vector-like fermions introduced in the lepton sector, 
and $M_G$ the 
central value for the masses of the family gauge bosons. 
Taking $g_{F} \sim O(1)$, and, for the masses, the values required in the 
lepton sector to have a proper value for the neutrino of the $4^{th}$ 
generation, $\tan{\b}$ turns out to be much smaller than unity.
This makes the contribution to ${\cal M}$ due to the pseudo NG bosons 
negligible, being suppresed by the factor $\sin^2{\b}$. In the following we 
will give only the expressions for the contributions due to the NG bosons 
$\tilde{\Omega}_i^{\prime}$. 
All the expressions that will be given for the effective vertices will have 
to be multiplied by the factor $\cos^2{\b}$.
We will focus on the expressions for the effective vertices in the transition 
$s \rightarrow d$. 

\section{ The Z effective vertex }

To calculate the Z effective vertex we have to sum up all the contributions 
coming from the different vector-like fermions propagating in the loop diagrams. 
We will present separately the expressions for the amplitude due respectively 
to the $ F_{d} $ and ${\cal M}_1$ vector-like fermions.
We remark here that $ F_{d} $ is the down component of a doublet of 
$ SU(2)_L $ while  ${\cal M}_1$ is a singlet of $ SU(2)_L $ ( see Table 1). 
Because the dominant terms of the amplitudes are proportional to 
$ T_{3L} $ of the vector-like fermions, the contribution due to $F_d$ 
is singled out. 
The contribution from 
${\cal M}_1$ will be  suppressed by the factor $ 1/M_G^2 $ with $ M_G $ 
being the scale of breaking of the family symmetry.   
The amplitude due to $ F_{d} $ is as follows

\begin{eqnarray}     
{\cal M}_{Z}^{F,\mu} & = & 
\frac{-ig}{16{\pi}^2 \cos({\theta}_W)}g_{F}^2
\sum_{j=1}^3 A^{T}_{{\Omega},2j}A_{{\Omega},j1}   
2(g_{-}-g_{+})\left[ \frac{1}{2}\ln{\frac{M_{\Omega_j}^2}{M_F^2}} + 
\frac{1}{2}\frac{M_F^2}{M_F^2-M_{\Omega_j}^2} + \right. \nonumber \\
             &   &
\left. \frac{1}{2}\left(\frac{M_F^2}{M_F^2-M_{\Omega_j}^2}\right)^2
\ln{\frac{M_{\Omega_j}^2}{M_F^2}} - 
\frac{M_F^2}{M_F^2-M_{\Omega_j}^2}\ln{\frac{M_{\Omega_j}^2}{M_F^2}} \right]   
\bar{s}{\gamma}^{\m}(1-\g^5)d  \, ,
\end{eqnarray}    

\noindent
where 

\begin{equation}
g_{-}-g_{+} = - g_A = -T_{3L} = \frac{1}{2} \, , 
\end{equation}

\noindent
It is important to notice here that with $M_{\Omega_j}$ we have indicated the 
poles in the propagators of the longitudinal components of the gauge boson 
mass eigenstates $\tilde{\Omega}_i^{\prime}$.

Using for $M_{\Omega_j}$ the eigenvalues of the family gauge boson mass matrix 
in the simplistic form introduced for the lepton sector 

\begin{equation}
{\cal M}_{G}^2 = M_G^2\left(\begin{array}{cccc}
 1 &  b & 0 \\
 b &  1 & 0 \\
 0 &  0 & 1 
\end{array}\right) \, ,
\end{equation}

\noindent
given by

\begin{equation}
\begin{array}{lcl}
M_{\Omega_1}^2 & = & M_G^2\,(1+b) \, ,\\
M_{\Omega_2}^2 & = & M_G^2\,(1-b) \, ,\\
M_{\Omega_3}^2 & = & M_G^2 \, ,\\
\end{array}
\label{massomega}
\end{equation}

\noindent
where $b$ is a small parameter less than unity, and the orthogonal matrix 
$A_{\Omega}$ is given in Eq. \ref{matrixA}, we obtain the following final 
expression for $ {\cal M}_{Z}^{F,\mu} $

\begin{equation}
{\cal M}_{Z}^{F,\m}  = 
\frac{-ig}{16{\pi}^2 \cos({\theta}_W)}g_{F}^2\, C_{0}(x_F,b)\,
\bar{s}{\gamma}^{\m}(1-\g^5)d \, , 
\end{equation}

\noindent
with

\begin{eqnarray}
C_{0}(x_F,b) & = & \frac{1}{4\, {\left(1 + b - x_F \right)}^2\, 
{\left(-1 + b + x_F \right)}^2}  
\left[{-\left(-1 + b \right)}^2\, {\left(1 + b - x_F \right)}^2\, 
\ln (1 - b)  \right.  \nonumber \\
& & 
    \left. +  {\left(1 + b \right)}^2\, {\left(-1 + b + x_F \right)}^2\, 
\ln (1 + b) \right.  \nonumber \\
& &
\left.
- 2\, b\, x_F\, \left(b^2 - {\left(-1 + x_F \right)}^2 + 
          2\, \left(-1 + b^2 + x_F \right)\, \ln (x_F) \right) \right] \, ,  
\nonumber \\
\end{eqnarray}

\noindent
where $x_F$ is defined as

\begin{equation}
x_F = \frac{M_F^2}{M_G^2} \, .
\end{equation}

\noindent
Making the Taylor expansion of $C_{0}(x_F,b)$  in the parameter $b$ we found 
that the first non zero contribution comes from the linear term in $b$. 
So we can rewrite $C_{0}(x_F,b)$ as 

\begin{equation}
C_{0}(x_F,b) = b\, C_0(x_F) + {\cal O}(b^2) \, , 
\end{equation}

\noindent
with $C_{0}(x_F)$ given by

\begin{equation}
C_{0}(x_F) = \frac{1}{2}\left[
\frac{1+x_F}{(1-x_F)^2} + 
\frac{2x_F}{(1-x_F)^3}\ln{x_F} \right]  \, .
\label{C0}
\end{equation}

\noindent 
$ {\cal M}_{Z}^{F,\m} $ now takes the simple form

\begin{equation}     
{\cal M}_{Z}^{F,\m}  = 
\frac{-ig}{16{\pi}^2 \cos({\theta}_W)}g_{F}^2\,b\, C_{0}(x_F)\,
\bar{s}{\gamma}^{\m}(1-\g^5)d + {\cal O}(b^2) \, , 
\label{ampZF}
\end{equation}

\noindent
which shows, in the first approximation, an explicit linear dependence on the 
parameter $b$. 
In Appendix B, it will be shown in a simple example the mechanism that 
produces the $b$ dependence of the amplitude for FCNC processes and it will be 
compared with the GIM mechanism.

The contribution to the $Z$ effective vertex due to ${\cal M}_1$ takes the 
form

\begin{equation}     
{\cal M}_{Z}^{{\cal M}_1,\m}  = 
\frac{ig\sin^2({\theta}_W)}{16{\pi}^2\cos({\theta}_W)}
\frac{g_{F}^2}{4M_{{\cal M}_1}^2}\,D_{0}(x_M,b)\,
\bar{s}(q^2{\gamma}^{\m}-q^{\m}{\qs})(1+\g^5)d  \, ,
\end{equation}

\noindent
where $q$ is the momentum transfer and with

\begin{eqnarray}
D_{0}(x_M,b) & = & \left[       \frac{ b\,
              \left(-5\, b^4 + {\left(-1 + x_M \right)}^2\,
                  \left(5 + x_M \right)\,
                  \left(-1 + 7\, x_M \right)\right) }
{{\left(1 + b - x_M \right)}^3\, {\left(-1 + b + x_M \right)}^3} \right. 
\nonumber  \\
& & 
 \left.           + \frac{ 2\, b^3\, \left(5 + 
                        x_M\, \left(-22 + 23\, 
                          x_M \right) \right)  }
{{\left(1 + b - x_M \right)}^3\, {\left(-1 + b + x_M \right)}^3}  \right. 
\nonumber  \\
& &
    \left.          -\frac{3\, {\left(-1 + b \right)}^2\,
              \left(-1 + b + 3\, x_M \right)\,
              \ln (\frac{x_M}{1 - b})}{{\left(-1 + b + 
                        x_M \right)}^4} \right. 
\nonumber  \\
& &
   \left.
                    -\frac{3\, {\left(1 + b \right)}^2\,
              \left(1 + b - 3\, x_M \right)\,
              \ln (\frac{x_M}{1 + b})}{{\left(1 + b - 
                        x_M \right)}^4}  \right] \frac{x_M}{27} \, , 
\label{D0}
\end{eqnarray}

\noindent
where $x_M$ is defined as

\begin{equation}
x_M = \frac{M_{{\cal M}_{1}}^2}{M_G^2} \, .
\end{equation}

\noindent
It is important to notice here that the structure of the operator in 
${\cal M}_{Z}^{{\cal M}_1,\m}$ is similar in character to that for the photon. 
In fact,  $q_{\mu}{\cal M}_{Z}^{{\cal M}_1,\m}=0$. This happens 
because ${\cal M}_1$ is a singlet of $SU(2)$ and so the current of 
${\cal M}_1$ that 
couples to $Z$ has the vectorial nature characteristic of the currents 
interacting with the photon. 
Making a Taylor expansion of the function $D_{0}(x_M,b)$ in the parameter $b$ 
we found also that in this case the first non vanishing term in the expansion 
is that one linear in $b$. So we can rewrite $D_{0}(x_M,b)$ as 

\begin{equation}
D_{0}(x_M,b) = b\,D_{0}(x_M) + {\cal O}(b^2) \, ,
\end{equation}

\noindent
with $D_{0}(x_M)$ given by

\begin{eqnarray}
D_{0}(x_M) & = & \frac{x_M}{27}\left[
\frac{11 + 6\ln{x_M} - 18x_M\ln{x_M} - 63x_M + 45x_M^2 + 7x_M^3 -  
36x_M^2\ln{x_M} }{(1-x_M)^5} \right]  \, . \nonumber \\
\end{eqnarray}

\noindent
Now ${\cal M}_{Z}^{{\cal M}_1,\m}$ takes the simple form

\begin{eqnarray}     
{\cal M}_{Z}^{{\cal M}_1,\m} & = & 
\frac{ig\sin^2({\theta}_W)}{16{\pi}^2\cos({\theta}_W)}
\frac{g_{F}^2}{4M_{M_1}^2}\,b\,D_{0}(x_M)\,
\bar{s}(q^2{\gamma}^{\m}-q^{\m}{\qs})(1+\g^5)d + {\cal O}(b^2) \, . 
\nonumber \\ 
\end{eqnarray}

For comparison we show the explicit expression of the corresponding 
contribution from 
the SM, as given in Ref. \cite{Buras1}.

\begin{eqnarray}     
{\cal M}_{Z}^{SM,\m} & = & 
\frac{ig}{16{\pi}^2 \cos({\theta}_W)}g^2 V_{ts}^{*}V_{td}C_{0}^{SM}(x_t)
\bar{s}{\gamma}^{\m}(1-\g^5)d \, ,
\label{ampZSM}
\end{eqnarray}    

\noindent
where

\begin{equation}
C_{0}^{SM}(x_t) = \frac{x_t}{8}\left[
\frac{x_t-6}{x_t-1} + 
\frac{3x_t+2}{(x_t-1)^2}\ln{x_t} \right] \, ,
\end{equation}

\noindent
with $x_t$ defined as

\begin{equation}
x_t=\frac{m_t^2}{M_W^2} \, ,
\end{equation}

\noindent
and $g$ is the $SU(2)_L$ gauge coupling.

\section{ The ${\gamma}$ effective vertex }

The effective vertex of $\gamma$ comes also from the contributions due 
the $F_d$ and ${\cal M}_1$ vector-like fermions. What is remarkable is that 
these two contributions will contain the {\em same function} $D_0$ 
(Eq. \ref{D0}), 
which appears in the contribution of $ {\cal M}_1 $ to the $Z$ penguin. This 
happens because, as mentioned before, ${\cal M}_1$ is a singlet of $SU(2)$ 
and is coupled to $Z$ in the same way as to $\gamma$ 
except for the coupling constant. 
The contributions to the effective $\gamma$ vertex from $F_d$ and 
${\cal M}_1$ will differ from each other in the form of the operators and 
with $x_F$ and $x_M$ appearing in $D_0$ respectively.  
We obtain the following final expressions

\begin{eqnarray}     
{\cal M}_{\gamma}^{F,\m} & = & 
\frac{-ie}{16{\pi}^2}\frac{g_{F}^2}{4M_F^2}\,b\,D_{0}(x_F) 
\bar{s}(q^2{\gamma}^{\m}-q^{\m}{\qs})(1-\g^5)d + {\cal O}(b^2) \, , 
\end{eqnarray}

\noindent
and 

\begin{eqnarray}     
{\cal M}_{\gamma}^{{\cal M}_1,\m} & = & 
\frac{-ie}{16{\pi}^2}\frac{g_{F}^2}{4M_{{\cal M}_1}^2}\,b\,D_{0}(x_M)
\bar{s}(q^2{\gamma}^{\m}-q^{\m}{\qs})(1+\g^5)d + {\cal O}(b^2) \, .
\end{eqnarray}

The corresponding expression for the SM as given in Ref. \cite{Buras1} is

\begin{eqnarray}     
{\cal M}_{\gamma}^{SM,\m} & = & 
\frac{-ie}{16{\pi}^2}\frac{g^2}{4M_W^2}V_{ts}^{*}V_{td}\,D_{0}^{SM}(x_t)\,
\bar{s}(q^2{\gamma}^{\m}-q^{\m}{\qs})(1-\g^5)d  \, , 
\end{eqnarray}    

\noindent
with

\begin{eqnarray}
D_{0}^{SM}(x_t) & = & -\frac{4}{9}\ln{x_t} + \frac{-19x_t^3 + 25x_t^2}
{36(x_t-1)^3} + \frac{x_t^2(5x_t^2-2x_t-6)}{18(x_t-1)^4}\ln{x_t} \, .
\end{eqnarray}

\section{ The gluon effective vertex }

The function $D_0$ which appears in the $\gamma$ vertex also appears in 
the gluon effective vertex,
with the only difference appearing in the prefactor due to the different 
nature of the charge. This is so because $g$ and $\gamma$ have vector 
couplings to the same particles in our model. This does not 
happen in the SM where $\gamma$ can couple to  
$W$ while the gluon cannot. As a result, the number of 
diagrams in the SM, used to describe the effective vertices, and the 
functions that appear in there, are different for the gluon and for 
$\gamma$.
We introduce the function $E_0$ given by

\begin{equation}
E_0 = -3D_0 \, ,
\end{equation}

\noindent
to absorb the factor $-1/3$ in $D_0$ coming from the electric charge 
of the quarks. 
The final expressions for the contributions to the gluon effective vertex 
due to $F_d$ and ${\cal M}_1$ are ($\alpha$, $\beta$ are color indices and $g_s$ is the $SU(3)_c$ coupling)

\begin{eqnarray}     
{\cal M}_{g}^{F,a,\m} & = & 
\frac{i{g_s}}{16{\pi}^2}\frac{g_{F}^2}{4M_F^2}bE_{0}(x_F)
\bar{s}_{\a}(q^2{\gamma}^{\m}-q^{\m}{\qs})(1-\g^5)T_{\a\b}^{a}d_{\b} + 
{\cal O}(b^2) 
\, ,
\end{eqnarray}

\noindent
and 

\begin{eqnarray}     
{\cal M}_{g}^{{\cal M}_1,a,\m} & = & 
\frac{i{g_s}}{16{\pi}^2}\frac{g_{F}^2}{4M_{{\cal M}_1}^2}bE_{0}(x_M)
\bar{s}_{\a}(q^2{\gamma}^{\m}-q^{\m}{\qs})(1+\g^5)T_{\a\b}^{a}d_{\b} + 
{\cal O}(b^2) \, ,
\end{eqnarray}    

\noindent
where $(T_{\a\b})^{a}$ is the $a$-th generator of $SU(3)_c$.
The corresponding expression for the SM as given in Ref. \cite{Buras1} is

\begin{eqnarray}     
{\cal M}_{g,SM}^{a,\m} & = & 
\frac{i{g_s}}{16{\pi}^2}\frac{g^2}{4M_W^2}V_{ts}^{*}V_{td}E_{0}^{SM}(x_t)
\bar{s}_{\a}(q^2{\gamma}^{\m}-q^{\m}{\qs})(1-\g^5)T_{\a\b}^{a}d_{\b} \, ,
\end{eqnarray}    

\noindent
with

\begin{eqnarray}
E_{0}^{SM}(x_t) & = & -\frac{2}{3}\ln{x_t} + \frac{x_t^2(15-16x_t + 
4x_t^2)}{6(1-x_t)^4}\ln{x_t} + \frac{x_t(18 - 11x_t-x_t^2)}{12(1-x_t)^3} 
\, .
\end{eqnarray}

\noindent
Knowing the expressions for the effective vertices in the new model is what 
we need to derive completely the new contributions to the FCNC processes 
with ${\Delta}S=1$ involving penguin diagrams. These new contributions differ 
from  the corresponding ones in the SM only in the effective 
vertex. 

\section{Box amplitudes}

For the contributions to the FCNC processes with ${\Delta}S=1$ coming from 
box diagrams we have to calculate directly the amplitude. 
In Fig. 5 we show the new one loop diagrams describing FCNC processes with 
${\Delta}S=1$ coming from our model. 
The new contributions coming from the box diagrams to the FCNC processes 
with ${\Delta}S=1$ are suppresed by the factor $1/M_G^2$. This makes these 
box contributions negligible with respect to the dominant contribution, 
due to the $Z$ penguin, which is suppressed only by 
the factor $1/M_W^2$ coming from the Z propagator. 
The number of box diagrams is eight, considering only the ones in 
which the NG bosons 
propagating in the loop are the $\tilde{\Omega}_i^{\prime}$. 
The diagrams in which the pseudo NG bosons 
$Re\tilde{\rho}_i^{\prime}$ are propagating in the loop are suppressed by 
$\sin^4{\beta}$ while those which contain both NG bosons and pseudo NG 
bosons are suppressed by $\sin^2{\beta}$.
The first two diagrams (Fig. 5a,b) have the F vector-like particles propagating 
in the loop.
Neglecting terms of order $m_q^2/M_G^2$ with $m_q$ being the mass of the 
external 
quarks, we find the same amplitude for the two diagrams. We also use the  
identity

\begin{eqnarray}
& & \bar{s}(p_1){\gamma}^{\m}(1-\g^5)d(p_2) 
\bar{d}(p_4){\gamma}_{\m}(1-\g^5)d(p_3)  =  \nonumber \\
& & \bar{s}(p_1){\gamma}^{\m}(1-\g^5)d(p_3) 
\bar{d}(p_4){\gamma}^{\m}(1-\g^5)d(p_2) \, ,
\end{eqnarray}  

\noindent
applying Fierz transformations.
For both diagrams the amplitude is given by

\begin{eqnarray}
{\cal M}_{box} &  =  &
\frac{i g_{F}^4}{16 \pi^2 M_G^2}\,B_0(x_F,b)\,
\bar{s}{\gamma}^{\m}(1-\g^5)d \,\bar{d}{\gamma}_{\m}(1-\g^5)d \, , 
\label{ampbox}
\end{eqnarray}

\noindent
where $B_0(x_F,b)$ is given by

\begin{eqnarray}
B_0(x_F,b) & = &
  \frac{-{\left( 1 + b \right) }^2 + x^2 + 
     2\,\left( 1 + b \right) \,x\,
      \left( \ln (1 + b) - \ln (x) \right) }{2\,
     {\left( 1 + b - x \right) }^3} -  \nonumber  \\
& &
  \frac{ -{\left( 1 - b \right) }^2 + x^2 + 
       2\,\left( 1 - b \right) \,x\,
        \left( \ln (1 - b) - \ln (x) \right)  }
     {2\,{\left( 1 - b - x \right) }^3}  \, .
\end{eqnarray}

For the function $B_0(x_F,b)$, in the same way as for $C_0(x_F,b)$ and 
$D_0(x_F,b)$, we make a Taylor expansion in the parameter $b$. 
Also in this case as in the case for the $Z$ and $\gamma$ vertices the 
zeroth order term 
in the expansion is not present and the first non vanishing term is that 
one which is linear in $b$.   
The amplitude in Eq. \ref{ampbox} now takes the simple form 

\begin{eqnarray}
{\cal M}_{box} & = & 
\frac{i g_{F}^4}{16 \pi^2 M_G^2}\,b\,B_0(x_F)\,
\bar{s}{\gamma}^{\m}(1-\g^5)d \,\bar{d}{\gamma}^{\m}(1-\g^5)d + 
{\cal O}(b^2) \, , 
\end{eqnarray}

\noindent
where $B_0(x_F)$ is given by

\begin{eqnarray}
B_0(x_F) & = &  \frac{ 1 + \left( 4 - 5\,x \right) \,x + 
       2\,x\,\left( 2 + x \right) \,\ln (x) }
     {{\left( -1 + x \right) }^4} \, .
\end{eqnarray}

The amplitudes for the box diagrams in which we substitute the 
vector-like fermions F with ${{\cal M}_1}$ have similar expressions except 
in the 
operator which now has the form $(V+A)(V+A)$ instead of $(V-A)(V-A)$ and with
$x_M$ now appearing in the function $B_0$.  
As in Ref. \cite{PQ}, we expect $M_{{\cal M}_1}$to be much larger than 
$M_F$ so that $x_M=M_{{\cal M}_1}^2/M_G^2$ is closer to unity while 
$x_F=M_{F}^2/M_G^2$ can be much less than unity. As can be seen from Fig. 6,
$B_0$ varies by at most an order of magnitude for $0 < x_{F,M} \leq 1$. As 
a result we expect the contributions of the two diagrams due to ${\cal M}_1$
to be smaller than the corresponding ones due to F by not more than one 
order of magnitude. 
For the four box diagrams in which there are one vector-like particle 
of type F and one 
of type ${\cal M}_1$, the analytical expressions are too complicated to be 
written down in this paper. 
Instead we perform a numerical evaluation for the function 
$B_{mix 0}(x_F,x_M)$ in 
the mixed case giving us values between $B_0(x_F)$ and 
$B_0(x_M)$.
The operators for these last four box diagrams are of the form $(V-A)(V+A)$ 
and $(V+A)(V-A)$.

The amplitude for the box diagrams in the SM \cite{Buras1} is given by

\begin{equation}
{\cal M}_{box}^{SM}  =  
\frac{g^4}{64 \pi^2 M_W^2}\,V_{ts}^{*}V_{td}|V_{td}|^2\,B_{0}^{SM}(x_t)\,
\bar{s}{\gamma}^{\m}(1-\g^5)d \,\bar{d}{\gamma}^{\m}(1-\g^5)d  \, , 
\end{equation}

\noindent
where

\begin{equation}
B_{0}^{SM}(x_t) = \frac{1}{4}\left[\frac{x_t}{1-x_t} + 
\frac{x_t\ln{x_t}}{(x_t-1)^2}\right] \, .
\end{equation}

\section{Constraint on $\lowercase{b}$ from the upper bound on 
$K_L \rightarrow \m^+ \m^- $ }

From our analysis of the new contributions to FCNC processes with 
${\Delta}S=1$, it turns out that the main contribution is given by the $Z$ 
penguin, all the other contributions being $1/M_G^2$ suppressed. 

The FCNC processes with ${\Delta}S=1$ in kaon physics are reasonably well 
described by the SM, with uncertainties coming from non perturbative QCD 
effects. On the other hand the apparent discrepancy for example between the 
SM estimates and the data invites for speculations about non standard 
contributions to $\epsilon^{\prime}/\epsilon$ \cite{Buras5}, and also for 
FCNC rare deacys there are still margins for effects of new physics 
\cite{Isidori1}.      
Now in our extension of the SM, the only non negligible 
contribution comes from the $Z$ penguin and corresponds to the operator $V-A$,
the same as in the $Z$ penguin of the SM. 

The way we proceed follows  Ref. \cite{Buras2}, \cite{Silvestrini}. 
First of all we write down 
the effective Lagrangians corresponding to the flavour-changing coupling 
of the $Z$ boson to down-type quarks in the SM and in our model, looking 
only, for the latter case, at the dominant contribution.
We have 

\begin{eqnarray}     
{\cal L}_{Z}^{SM} & = & 
\frac{g}{16{\pi}^2 \cos({\theta}_W)}g^2 Z_{ds}^{SM}
\bar{s}{\gamma}^{\m}(1-\g^5)d \, Z_{\m} + h.c. \, ,
\label{lagZSM}
\end{eqnarray}    

\noindent
with 

\begin{equation}
Z_{ds}^{SM} = V_{ts}^{*}V_{td}C_{0}^{SM}(x_t) \, ,
\label{ZSM}
\end{equation}

\noindent
and

\begin{equation}
{\cal L}_{Z}^{F}  = 
\frac{-g}{16{\pi}^2 \cos({\theta}_W)}g_{F}^2\,Z_{ds}^{F}
\bar{s}{\gamma}^{\m}(1-\g^5)d \,Z_{\m} + h.c. + {\cal O}(b^2) \, , 
\label{lagZF}
\end{equation}

\noindent
with

\begin{equation}
Z_{ds}^{F} = b\, C_{0}(x_F) \, .
\label{ZF}
\end{equation}

We  notice here that  Eq. \ref{lagZSM} and \ref{lagZF} are related to  
Eq. \ref{ampZSM} and \ref{ampZF} through the equation

\begin{equation}
{\cal L} = -i {\cal M}^{\m}Z_{\m} + h.c. \, .
\end{equation}

\noindent
As showed in Ref. \cite{Buras2} the coupling $Z_{ds}^{SM}$ is complex, 
being the product of the quantity $V_{ts}^{*}V_{td}$, which is complex, 
where $V_{ij}$ are elements of the CKM matrix, and 
the function $C_{0}(x_t)$ which is real. In particular, 
from the standard analysis of the unitary triangle, one has

\begin{eqnarray}
Im(V_{ts}^{*}V_{td}) & = & (1.38 \pm 0.33)\cdot10^{-4} \, , \\
Re(V_{ts}^{*}V_{td}) & = & -(3.2 \pm 0.9)\cdot10^{-4} \, ,
\label{Relambdat}
\end{eqnarray}

\noindent
which, using for $C_0(x_t)$ the value of 0.79 corresponding to the 
central value of the top quark mass, $\bar{m}_t(m_t) = 166 \, GeV$, 
give

\begin{eqnarray}
ImZ_{ds}^{SM} & = & (1.09 \pm 0.26)\cdot10^{-4} \, , \\
ReZ_{ds}^{SM} & = & -(2.54 \pm 0.71)\cdot10^{-4} \, .
\label{ReZSM}
\end{eqnarray}
  
\noindent
The complex nature of the coupling $Z_{ds}^{SM}$ is responsible for CP 
violation in the SM, as can be verified by
looking at the expression for $\epsilon^{\prime}/\epsilon$ as given in Ref. 
\cite{Buras2}: 

\begin{equation}
\frac{\epsilon^{\prime}}{\epsilon} = 
\left(\frac{\epsilon^{\prime}}{\epsilon}\right)_Z + 
\left(\frac{\epsilon^{\prime}}{\epsilon}\right)_{Rest} \, ,
\end{equation}

\noindent
where

\begin{equation}
\left(\frac{\epsilon^{\prime}}{\epsilon} \right)_Z = 
ImZ_{ds}\left[1.2-R_s|r_Z^{(8)}|
B_8^{(3/2)}\right] \, ,
\label{epsilonZ}
\end{equation}

\noindent
and
 
\begin{equation}
\left(\frac{\epsilon^{\prime}}{\epsilon} \right)_{Rest} = 
Im(V_{ts}^{*}V_{td})\left[-2.3 + R_s\left[ 1.1|r_Z^{(8)}|
B_6^{(1/2)} + (1.0+0.12|r_Z^{(8)}|)B_8^{(3/2)} \right] \right]\, ,
\end{equation}

\noindent
which is proportional to $Im(V_{ts}^{*}V_{td})$.
All the parameters appearing in the above expressions are fully described 
in Ref. \cite{Buras2}.
As pointed out in Ref. \cite{Buras2}, 
if we assume that no new operators in addition to those present in the SM 
contribute, and this is true in the approximation of neglecting the new 
contributions which are suppressed by the factor $1/M_G^2$, the replacement 
$Z_{ds}^{SM} \rightarrow Z_{ds}$, where 

\begin{equation}
Z_{ds} = Z_{ds}^{SM} + \frac{g_F^2}{g^2} Z_{ds}^{F} \, ,
\label{Z}
\end{equation}

\noindent
a relation that holds separately for $ReZ_{ds}$ and $ImZ_{ds}$

\begin{equation}
ReZ_{ds} = ReZ_{ds}^{SM} + \frac{g_F^2}{g^2} ReZ_{ds}^{F} \, ,
\end{equation}

\begin{equation}
ImZ_{ds} = ImZ_{ds}^{SM} + \frac{g_F^2}{g^2} ImZ_{ds}^{F} \, ,
\end{equation}

\noindent
which is justified without the modification of QCD renormalization group effects 
evaluated at NLO level for scales below ${\cal O}(m_t)$.
This means that to look at the effects of new physics, 
described by a modified effective coupling of the $Z$ boson to 
down-type quarks, to the quantity $\epsilon^{\prime}/\epsilon$, 
we just need to 
substitute $ImZ_{ds}$ to $ImZ_{ds}^{SM}$ in Eq. \ref{epsilonZ}. 
Looking at  Eq. \ref{ZF}, $Z_{ds}^F$ is the product of the parameter $b$, 
introduced in Eq. \ref{massomega}, 
which has been chosen real and the function $C_0(x_F)$, which  
is also real making $Z_{ds}^F$ real. 
This implies that in our model there are no corrections to 
$ImZ_{ds}^{SM}$ and consequently to $\epsilon^{\prime}/\epsilon$.  
Obviously this means that we cannot use the experimental results on 
$\epsilon^{\prime}/\epsilon$ to constrain the parameter $b$.
As shown in Ref. \cite{Buras2}, if new physics affects $ReZ_{ds}$, as it is 
the case in our model, the process to look at to reveal effects of new 
physics is the FCNC decay $K_L \rightarrow \m^+ \m^-$, whose experimental 
branching ratio $BR(K_L \rightarrow \m^+ \m^-)=(7.2 \pm 0.5)\cdot10^{-9}$ 
\cite{Heinson1}. 
The effects of new physics appear in the short-distance (SD) contribution to   
$BR(K_L \rightarrow \m^+ \m^-)$

\begin{equation}
BR(K_L \rightarrow \m^+ \m^-)_{SD} = 6.32\cdot10^{-3} 
\left[ Re Z_{ds} - B_0 Re(V_{ts}^{*}V_{td}) + \bar{\Delta}_c \right]^2 \, ,
\end{equation}

\noindent 
where $B_0=-0.182$ is the box diagram function evaluated at 
$\bar{m}_t(m_t) = 166 GeV$, and 

\begin{equation}
\bar{\Delta}_c=-(6.54 \pm 0.60)\cdot10^{-5} \, ,
\end{equation}

\noindent
represents the charm contribution \cite{Buchalla3}.
From the analysis in Ref. \cite{Buras2} of the long distance (LD) and 
SD contributions to $BR(K_L \rightarrow \m^+ \m^-)$ the highest possible 
value for $BR(K_L \rightarrow \m^+ \m^-)_{SD}$ is derived 

\begin{equation}
BR(K_L \rightarrow \m^+ \m^-)_{SD} < 2.8 \cdot 10^{-9}  \, .
\label{BRSD}
\end{equation} 

\noindent
Now using Eq. \ref{BRSD} together with Eq. \ref{Z}, \ref{ReZSM} and \ref{ZF} 
we obtain the following upper bound on the parameter $b$

\begin{equation}
b < 2.5\cdot10^{-4} \, ,
\label{bconstr}
\end{equation}

\noindent
where in $Z_{ds}^F$ we have used $C_0(10^{-6})\simeq 0.5$ which corresponds to 
$M_F=200GeV$ and $M_G=200TeV$, and in $Z_{ds}$ we have used $g=0.65$ and 
$g_F=1$. 

Looking at the expressions for $BR(K_L \rightarrow \pi ^0 \n \bar{\n})$, 
$BR(K_L \rightarrow \pi ^0 e^+ e^-)$ and 
$BR(K^+ \rightarrow \pi ^+ \n \bar{\n})$ given in Ref. \cite{Buras2}, we 
observe that there are no new contributions from our model to the first two decays, 
which depend only on $ImZ_{ds}$,
while there is a contribution to 
$K^+ \rightarrow \pi ^+ \n \bar{\n}$, which depends also on $ReZ_{ds}$. 
In Ref. \cite{Buras2}, the expression for the upper bound on 
$BR(K^+ \rightarrow \pi ^+ \n \bar{\n})$ is given in terms of 
$BR(K_L \rightarrow \pi ^0 \n \bar{\n})$, and the parameter $\kappa$ related to 
$BR(K_L \rightarrow \m^+ \m^-)_{SD}$ by the relation

\begin{equation}
BR(K_L \rightarrow \m^+ \m^-)_{SD} = \kappa \cdot10^{-9} \, .
\end{equation}

\noindent
Now, if as in Ref. \cite{Buras2} we use the upper bound for 
$BR(K_L \rightarrow \m^+ \m^-)_{SD}$ given in Eq. \ref{BRSD}, we obtain

\begin{equation} 
BR(K^+ \rightarrow \pi ^+ \n \bar{\n}) < 0.229 \cdot 
BR(K_L \rightarrow \pi ^0 \n \bar{\n}) + 1.76\cdot 10^{-10} \, ,
\end{equation}

\noindent
while, if we use the upper theoretical value for the 
$BR(K_L \rightarrow \m^+ \m^-)_{SD}^{SM}$ given in Ref. \cite{Buras2}, 
we obtain 

\begin{equation} 
BR(K^+ \rightarrow \pi ^+ \n \bar{\n}) < 0.229 \cdot 
BR(K_L \rightarrow \pi ^0 \n \bar{\n}) + 1.15\cdot 10^{-10} \, .
\end{equation}

\noindent
In Ref. \cite{Buras2}, for the particular scenario in which all the effects 
of new physics are encoded in the effective coupling $Z_{ds}$, 
$BR(K_L \rightarrow \pi ^0 \n \bar{\n})$ is estimated 
to range in the interval $(1.3\cdot10^{-10}, 2.4\cdot10^{-10})$ for 
$B_8^{(3/2)}=0.6$, and this makes the contribution due to the maximum value for 
$BR(K_L \rightarrow \m^+ \m^-)_{SD}$ in the expression for the upper bound of 
$BR(K^{+} \rightarrow \pi ^+ \n \bar{\n})$ the dominant contribution.  
The difference between the two upper bounds for 
$BR(K^{+} \rightarrow \pi ^+ \n \bar{\n})$
shows the room for a possible 
contribution of new physics associated to the quantity $ReZ_{ds}$.

Using the constraint on $b$ derived in Eq. \ref{bconstr}, 
we can now check if the amplitudes that 
we know to be $1/M_G^2$ suppressed are negligible with respect 
to the corresponding SM amplitudes. 
In the case of the photon, matching the absolute value of the functions 
which multiplies the $V-A$ operator for the new amplitude and the SM one, 
we have, after cancelling out all the common factors,

\begin{equation}
\frac{M_W^2}{M_F^2}
\left|\frac{g_{F}^2bD_0(x_F)}{g^2Re(V_{ts}^{*}V_{td})D_0^{SM}(x_t)}\right|
\leq 1.7\cdot10^{-6} \, ,
\end{equation}

\noindent
where we have used $D_0(10^{-6})\simeq -2.66\cdot10^{-6}$, and 
for the SM quantities $D_0^{SM}(4.27) \simeq -0.46$, $Re(V_{ts}^{*}V_{td})$ 
as given in Eq. \ref{Relambdat},  and obviously the 
constraint on $b$ from Eq. \ref{bconstr}.

In the case of the gluon we have    

\begin{equation}
\frac{M_W^2}{M_F^2}
\left|\frac{g_{F}^2bE_0(x_F)}{g^2 Re(V_{ts}^{*}V_{td})E_0^{SM}(x_t)}\right|
\leq 8.7\cdot10^{-6} \, ,
\end{equation}

\noindent
where we have used $E_0^{SM}(4.27)\simeq0.27$.
From what we have seen above, the amplitudes with the $V-A$ operator for the 
photon and for the gluon, which have corresponding ones in the SM, 
are negligible. 
Now the amplitudes with the $V+A$ operator which appears in our model, and 
are present in the effective vertices for the $Z$, the photon and the gluon, 
are also $1/M_G^2$ suppressed and are not relevant.

For the box diagrams, we look at the function which multiplies the 
$(V-A)(V-A)$ operator. We obtain the ratio

\begin{equation}
\frac{M_W^2}{M_G^2}
\left|\frac{4g_{F}^4bB_0(x_F)}{g^4Re(V_{ts}^{*}V_{td})|V_{td}|^2
B_0^{SM}(x_t)}\right|
\leq 0.19 \, ,
\end{equation}

\noindent
where we have used $B_0(10^{-6})\simeq1$ and for the SM quantities 
$B_0^{SM}(4.27)\simeq-0.18$ and 
$|V_{td}| \simeq 9.1 \cdot 10^{-3}$. Similarly, 
the amplitudes containing the operators $(V+A)(V+A)$, 
$(V-A)(V+A)$ and  $(V+A)(V-A)$ are also not relevant.
It is important to notice here that, for the box diagrams, the $1/M_G^2$ 
suppression of the new contribution is balanced  
by the $|V_{td}|^2$ suppression of the SM one. 

In Ref. \cite{PQ} it has been shown in two numerical examples how it is 
possible to derive the neutrino mass splittings $\Delta m_{32}^2$ and 
$\Delta m_{21}^2$ for different values of $b$. In fact, in Ref. \cite{PQ}, 
$b$ has been introduced as a free parameter chosen to be smaller than unity.
In this paper it has been shown how it is possible to constrain the parameter 
$b$ by looking at the quark sector. 
In accordance with the upper bound for $b$ derived in this paper and 
presented in Eq. \ref{bconstr}, we give here the neutrino mass splittings 
obtained numerically in Ref. \cite{PQ} for the particular choice of the 
parameter $b=0.000095$

\begin{equation}
\Delta m_{32}^2 = 2.02 \cdot 10^{-3} eV^2 \, ,
\label{masssplit1}
\end{equation}
\begin{equation}
\Delta m_{21}^2= 5.497 \cdot 10^{-6} eV^2 \, .
\label{masssplit2}
\end{equation}     

It is important to notice here that, when one chooses a particular value 
of the parameter $b$, one also has to choose the internal loop variables
which appear in the function given in Eq. \ref{deltaI},
in order to reproduce the mass splittings of Eq. \ref{masssplit1} and 
\ref{masssplit2}. 
This means, as one can see from the two numerical examples with 
different values of the parameter $b$ in Ref. \cite{PQ}, that different  
values of $b$ will correspond to different sets of masses for the three 
light neutrinos.  

\section{Other Rare Kaon Decays beyond the SM}

Being sensitive to flavour dynamics from few MeV up to several TeV, rare 
kaon decays provide a powerful tool to test the SM and to search for new 
physics.
Decays like $K_L \rightarrow \mu e$ and $K_L \rightarrow \pi \mu e$  are 
completely forbidden  within the SM \cite{Isidori1}, 
where lepton flavour is conserved, and 
are also absolutely negligible if we simply extend the model by including 
Dirac type neutrino masses with the standard Yukawa mass term. 

In our model the neutrinos are still only Dirac, but the way they get masses 
is through loop diagrams, and processes like $K_L \rightarrow \mu e$ or  
$K_L \rightarrow \pi \mu e$ are made possible through the exchange of 
virtual NG bosons $\tilde{\Omega}_i^{\prime}$. They, in fact, as we have 
already said, can couple to different flavours and, in the way we build our 
model, are the same for the lepton and the quark sectors. 

In the following, we will calculate explicitly the branching ratio for the decay 
$K_L \rightarrow \mu e$ in our model, and we will give a theoretical range of 
values corresponding to different values for the mass of the vector-like 
fermions $M_{{\cal M}_1^{l,q}}$. 

This decay happens through five diagrams, shown in Fig. 8: four box diagrams 
that have to be considered with their corresponding crossed diagrams, and 
the diagram 
obtained by linking two effective $Z$ verteces by a virtual $Z$.  
It can be shown that the crossed diagrams are $b$ suppressed with 
respect to the ones shown in Fig. 8. The diagram where the virtual $Z$ is 
exchanged is also $b$ suppressed, 
but not $1/M_G^2$ suppressed as the surviving box diagrams and so we will 
expect the sum of all the contributions will depend on the values of the 
parameters $b$ and $M_G$.
The final amplitude for the process $K_L \rightarrow \mu e$ is given by

\begin{eqnarray}
{\cal M} & = & 
\frac{i g_{F}^2}{2M_G^2}\left[ 
\left( \frac{g_{F}^2}{4 \pi^2}B(x_{F_q},x_{F_l},b)
-\frac{g^2 g_{F}^2 M_G^2}{ 32\pi^4 M_Z^2 \cos^2{({\theta}_W})}b^2 
C_0(x_{F_q})C_0(x_{F_l}) \right)
\bar{s}{\gamma}^{\m}\,P_L\,d \,\bar{e}{\gamma}_{\m}\,P_L\,\mu \right.  
\nonumber  \\
& &  \left.
+ \frac{g_{F}^2}{4 \pi^2}B(x_{F_q},x_{M_l},b) 
\bar{s}{\gamma}^{\m}\,P_L\,d \,\bar{e}{\gamma}_{\m}\,P_R\,\mu \right.   
\nonumber  \\ 
& & \left.
+ \frac{g_{F}^2}{4 \pi^2}B(x_{M_q},x_{F_l},b) 
\bar{s}{\gamma}^{\m}\,P_R\,d \,\bar{e}{\gamma}_{\m}\,P_L\,\mu \right.   
\nonumber  \\ 
& & \left.
+ \frac{g_{F}^2}{4 \pi^2}B(x_{M_q},x_{M_l},b) 
\bar{s}{\gamma}^{\m}\,P_R\,d \,\bar{e}{\gamma}_{\m}\,P_R\,\mu \right] \, ,   
\label{ampk}
\end{eqnarray}

\noindent 
with $B$ being the function coming from the box diagrams (which is too
long to be written down explicitely in this 
paper), 
$C_0$ given in Eq. \ref{C0}, and $P_{L,R}=\frac{1\mp\gamma^5}{2}$.
To obtain the $BR$ we can use the expression given in ref. \cite{Rizzo} for 
our case

\begin{equation}
BR  =  11.24\cdot 10^{-12}\left[\frac{g_{F}}{g}\frac{100TeV}{M_G}\right]^4
(C_{Lq}-C_{Rq})^2(C_{Ll}^2+C_{Rl}^2) \, ,
\label{BRKLmue}
\end{equation}

\noindent
that for our purpose can be rewritten as: 

\begin{eqnarray}
BR & = & 11.24\cdot 10^{-12}\left[\frac{g_{F}}{g}\frac{100TeV}{M_G}\right]^4
[(C_{Lq}C_{Ll})^2 + (C_{Lq}C_{Rl})^2 + (C_{Rq}C_{Ll})^2 + (C_{Rq}C_{Rl})^2 
\nonumber   \\
& & - 2(C_{Lq}C_{Ll})(C_{Rq}C_{Ll})  - 2(C_{Lq}C_{Rl})(C_{Rq}C_{Rl})] \, ,
\label{BR}
\end{eqnarray}

\noindent
where $C_{Lq},C_{Rq},C_{Ll},C_{Rl}$ appear in the expression for the 
operator ${\cal O}_{V,A}$ 

\begin{equation}
{\cal O}_{V,A}  =  \frac{g_F^2}{2M_G^2}\bar{s}{\gamma}^{\m}
[C_{Lq}P_L + C_{Rq}P_R]d \,\bar{e}{\gamma}_{\m}[C_{Ll}P_L + C_{Rl}P_R]\mu 
+ h.c. \, .
\end{equation}

\noindent
From the way we have written the amplitude ${\cal M}$ in Eq. \ref{ampk} it is 
possible to isolate the four contributions $C_{Lq}C_{Ll}$, $C_{Lq}C_{Rl}$, 
$C_{Rq}C_{Ll}$ and $C_{Rq}C_{Rl}$ that appear in Eq. \ref{BR}.

\begin{eqnarray}
C_{Lq}C_{Ll} & = &  \frac{g_{F}^2}{4 \pi^2}B(x_{F_q},x_{F_l},b)
-\frac{g^2 g_{F}^2 M_G^2}{ 32\pi^4 M_Z^2 \cos^2{({\theta}_W})}b^2 
C_0(x_{F_q})C_0(x_{F_l}) \, , \nonumber  \\
C_{Lq}C_{Rl} & = &  \frac{g_{F}^2}{4 \pi^2}B(x_{F_q},x_{M_l},b) \, , \\
C_{Rq}C_{Ll} & = &  \frac{g_{F}^2}{4 \pi^2}B(x_{M_q},x_{F_l},b) \, , 
\nonumber \\
C_{Rq}C_{Rl} & = &  \frac{g_{F}^2}{4 \pi^2}B(x_{M_q},x_{M_l},b) \, .
\nonumber
\end{eqnarray}

In Fig. 9, the logarithm of the quantity 
$(C_{Lq}-C_{Rq})^2(C_{Ll}^2+C_{Rl}^2)$ is plotted 
as a function of the logarithm of the ratio $x_M/x_F$, 
with $M_G=200TeV$, $x_F=10^{-6}$ 
and using 
for $b$ the upper bound given in Eq. \ref{bconstr}, so that the contribution 
coming from the diagram with the virtual $Z$ has its highest value. 
In doing the plot of Fig. 9 we have also chosen 
$M_{{\cal M}^l_1}=M_{{\cal M}^q_1}$ and  $M_{F^l}=M_{F^q}=200GeV$.
In Fig. 10 the logarithm of $BR(K_L \rightarrow \mu e)$ is plotted
as a function of the logarithm of the ratio $x_M/x_F$, for the same values of 
the quantities $M_G$, $x_F$ and $b$ as in Fig. 9, and the same choice 
$M_{{\cal M}^l_1}=M_{{\cal M}^q_1}$ and  $M_{F^l}=M_{F^q}=200GeV$. 
It can be seen from Fig. 10 that  
$BR(K_L \rightarrow \mu e)$ depends strongly on $x_M$, starting from 
$6.65\cdot10^{-22}$ for $x_M/x_F=1$, and reaching  the asymptotic value 
of $1.02\cdot10^{-14}$.
It is important to notice here that the lowest value for 
$BR(K_L \rightarrow \mu e)$, when $x_M/x_F=1$, corresponds to the situation 
in which the contributions of the box diagrams cancel out and the only 
contribution left is due to the diagram with the virtual $Z$. 
In this case we have   

\begin{eqnarray}
BR(K_L \rightarrow \mu e) 
& = & 11.24\cdot 10^{-12}\left[\frac{g_{F}}{g}100TeV\right]^4
\left(\frac{g^2 g_{F}^2}{ 32\pi^4 M_Z^2 \cos^2{({\theta}_W})}b^2 
C_0(x_{F_q})C_0(x_{F_l}) \right)^2 \nonumber \\ 
& = & 6.65\cdot10^{-22} \, ,
\end{eqnarray}

\noindent
where we have used for $b$ the upper bound given in Eq. \ref{bconstr}.
The asymptotic value for $BR(K_L \rightarrow \mu e)$, when $x_M/x_F \gg 1$, 
corresponds instead to the situation in which the box diagram with $F^l$ and 
$F^q$ propagating in the loop, gives the dominant contribution, and all the 
contributions from the other diagrams , including the one from the 
$Z$ penguin, are negligible. In this case we have

\begin{equation}
BR  =  11.24\cdot 10^{-12}\left[\frac{g_{F}}{g}\frac{100TeV}{M_G}\right]^4
(C_{Lq}C_{Ll})^2 = 1.02\cdot10^{-14} \, ,
\end{equation}

\noindent
which is independent of $b$.

The range of values for $BR(K_L \rightarrow \mu e)$ agrees with 
the actual experimental upper bound of $3\cdot10^{-12}$ at $90\, \% \,CL$ 
\cite{Jungmann}. This range is however quite wide, spanning eight
orders of magnitude, and is due to the fact that the branching ratio
has a strong dependence on the masses 
of the vector-like fermions $F^{l,q}$ and ${\cal M}_1^{l,q}$. 
Notice from Table 1 that $F^{l,q}$
are the vector-like leptons and quarks which are $SU(2)_L$ doublets,
while ${\cal M}_1^{l,q}$ denotes an $SU(2)_L$ singlet charged lepton
and quark respectively. It is interesting to note from Fig. 10 that,
in the case where all vector-like fermions are in the interesting
mass range  (a few hundreds of GeV's) where they can be produced 
by QCD or Electroweak processes,
$BR(K_L \rightarrow \mu e)$ is hopelessly small to have a chance to be
observed (the lower flat part of Fig. 10). At the other extreme (the
upper flat part of Fig. 10), $BR(K_L \rightarrow \mu e) \sim 10^{-14}$ 
while the mass of ${\cal M}_1^{l,q}$ is unreachable (of O(few hundreds
TeV's) assuming the mass of $F^{l,q}$ is of order a few hundreds of GeV's)
by any conceivable earthbound machine. 

\section{Epilogue}

We have presented a link between the quark sector and the lepton
sector based on a model of neutrino masses \cite{PQ}. This
link manifests itself in a common (small) parameter, $b$, which appears in 
both sectors. In the neutrino sector, $b$ is crucial in determining 
the magnitude of the mass splitting $\Delta m^2$, and, in an indirect
manner, the masses themselves. In fact, the model presented in Ref.
\cite{PQ} deals with the case of three light, degenerate
neutrinos, and the lifting of this degeneracy is proportional to $b$.
In the quark sector, as we have seen above,
this same parameter appears in various FCNC penguin operators
which are seen to be linear in $b$ at the lowest order. It is
well-known that these penguin operators are important in various
aspects of Kaon physics: $\epsilon^{\prime}/\epsilon$, rare K decays,
etc... Strong constraints in this sector would also constrain the
neutrino sector as well. It turns out, as we have shown above, that
there is no contribution to $\epsilon^{\prime}/\epsilon$ from our model,
to lowest order in $b$. However, at the same order in $b$, our
model makes a contribution to the rare decay process $K_{L} \rightarrow
\mu^{+} \mu^{-}$. Taking into account the upper bound on the short
distance contribution to that process, we found a strong constraint on
$b$ which is $b < 2.5 \times 10^{-4}$. 

We have also calculated the $BR(K_L \rightarrow \mu e)$, finding that it 
strongly depends on the masses of the vector-like fermions 
$F^{l,q}$ and ${\cal M}_1^{l,q}$ and on $b$ for a certain range
of those masses. We found that 
$BR(K_L \rightarrow \mu e)$ goes from $6.65\cdot10^{-22}$ which makes 
this decay practically unobservable, to $1.02\cdot10^{-14}$, with the choice 
of $M_G=200 TeV$ and $x_F=10^{-6}$, when $x_M/x_F$ goes from $1$ to $10^{12}$.
For $b$ we have used the upper bound derived from 
$K_{L} \rightarrow \mu^{+} \mu^{-}$. As one can see from Fig. 10, the
first case ($6.65\cdot10^{-22}$) corresponds to the interesting
situation where it might be possible to produce and observe these
vector-like particles since their masses could lie in the few-hundred
GeV region \cite{PQ2}, despite the fact that one will not be able to
observe $K_L \rightarrow \mu e$. The second case ($1.02\cdot10^{-14}$)
corresponds to a possible observation for $K_L \rightarrow \mu e$, while
forsaking that of the singlet quark and charged lepton ${\cal M}_1^{l,q}$.
And finally, there are these in-between cases as can be seen from Fig. 10. 

Because of the upper limit on $b$, and obviously because of the 
choice of heavy family gauge bosons, $M_G$ of ${\cal O}(100TeV)$,  
it turns out that the physics of the kaon sector by itself, in our
model, is not too different from the SM. 
Nevertheless, we have seen that there is still some margin for 
possible contributions of new physics to $K_{L} \rightarrow
\mu^{+} \mu^{-}$, the bound on $K^{+} \rightarrow \pi^{+} 
\nu \bar{\nu}$. As for the decay $K_L \rightarrow \mu e$, which
is forbidden in the SM, we have seen that our model can make
a non-negligible contribution ($BR \sim 10^{-14}$) which is
practically independent of $b$. In the region where it depends strongly
on $b$ (the lower flat region of Fig. 10), the branching ratio is
negligible, practically similar to the SM with a Dirac neutrino.
In that region, as we have stressed above, the new physics signal
would be the production and observation of the vector-like fermions.

The bound on $b$ could have interesting implications on the neutrino
sector itself, as we have mentioned above. It was shown in Ref.
\cite{PQ} how the parameter $b$ affects the mass splitting
of three formely degenerate neutrinos. In particular, it was shown how
$\Delta m^2$ is sensitive to $b$. However, it was also shown how
$b$ indirectly affects the overall magnitude of the masses. Unfortunately,
at the present time, one is quite far experimentally from a direct 
determination of the masses themselves. Needless to say, future
experiments are of paramount importance to this crucial question.

\acknowledgments

This work is supported in parts by the US Department
of Energy under grant No. DE-A505-89ER40518 (PQH),
and by the Italian INFN Fellowship (AS).

\section{Appendix A} 
In this appendix, we will summarize the results of Ref. \cite{PQ}.
The Yukawa part of the Lagrangian involving leptons can be
written as

\begin{eqnarray}
{\cal L}^Y_{Lepton}& =& g_E \bar{l}_L^{\alpha} \phi e_{\alpha\, R} +
G_1 \bar{l}^{\alpha}_{L} \Omega_{\alpha} F_{R} +
G_{{\cal M}_1} \bar{F}_{L} \phi {\cal M}_{1R}+
G_{{\cal M}_2} \bar{F}_{L} \tilde{\phi} {\cal M}_{2R} +
G_2 \bar{{\cal M}}_{1L} \Omega_{\alpha} e^{\alpha}_{R} + \nonumber \\
          &  &G_3 \bar{\cal{M}}_{2L} \rho^{\alpha}_{m} \eta^{m}_{\alpha R}
+ M_F \bar{F}_L F_R + M_{{\cal M}_1} \bar{{\cal M}}_{1L} {\cal M}_{1R} +
M_{{\cal M}_2} \bar{{\cal M}}_{2L} {\cal M}_{2R} + h.c. \, .
\label{lag1}
\end{eqnarray}

\noindent
which, after integrating out the vector-like fermions F, ${\cal M}_1$ and 
${\cal M}_2$, brings to the effective Lagrangian

\begin{eqnarray}
{\cal L}^{Y,eff}_{Lepton}& =& g_E \bar{l}_L^{\alpha} \phi e_{\alpha \,R} +
G_E \bar{l}^{\alpha}_{L} (\Omega_{\alpha} \phi \Omega^{\beta}) e_{\beta \,R} +
\nonumber \\
&  &G_N \bar{l}^{\alpha}_{L}(\Omega_{\alpha} \tilde{\phi} \rho^{\beta}_{i}) 
\eta^{i}_{\beta \,R} + h.c. ,
\end{eqnarray}

\noindent
where

\begin{equation}
G_E =\frac{G_1 G_{{\cal M}_1} G_2}{M_F M_{{\cal M}_1}};\, 
G_N =\frac{G_1 G_{{\cal M}_2} G_3}{M_F M_{{\cal M}_2}}.
\end{equation}

The main assumption in building the above Lagrangian is the conservation of 
lepton number L, thereby forbidding the presence of Majorana mass terms.
The way $<\Omega>=(0,0,0,V)$ and $<\rho>=(0,0,0,V^{\prime}\otimes s_1)$,
with $s_1 = \left( \begin{array}{c} 1 \\ 0 \end{array} \right)$, have been 
chosen, makes the neutrino of the $4^{th}$ family massive at tree 
level, 
while the others three neutrino remain massless. These three neutrinos 
would get a mass dynamically via loop diagrams.
For the neutrino of the $4^{th}$ family we have 

\begin{equation}
m_N = G_1 G_{{\cal M}_2} G_3 \frac{V\,V^{\prime}}{M_F\,M_{{\cal M}_2}} 
\frac{v}{\sqrt{2}} \, ,
\end{equation}

\noindent
which, for

\begin{equation}
V\, V^{\prime}/M_F\,M_{{\cal M}_2} \sim O(1) \, ,
\end{equation}

\noindent
can be expected to be even as heavy as $175 GeV$, satisfying the bound of 
$M_Z/2$ from LEP.

\noindent
For the three light neutrinos, one has 

\begin{equation}
m_\nu = m_N \frac{M_F M_{{\cal M}_2}}{V V^{\prime}}\frac{\sin(2\beta)}
{32\,\pi^2}\,\Delta I(G,P) \, ,
\end{equation}

\noindent
where $\Delta I(G,P)$ is given by

\begin{equation}
\Delta I(G,P) = \frac{1}{M_F-M_{{\cal M}_2}}
\{\frac{M_F[M_F^2(M_G^2\ln(\frac{M_G^2}{M_F^2})-M_P^2\ln
(\frac{M_P^2}{M_F^2}))+ 
M_G^2 M_P^2 \ln(\frac{M_P^2}{M_G^2})]}{(M_G^2-M_F^2)(M_P^2-M_F^2)} -(M_F 
\leftrightarrow M_{{\cal M}_2})\} \, .
\end{equation}

The main result that was obtained in Ref. \cite{PQ} 
is that one can obtain for $m_\nu$ 
a value of the ${\cal O}(eV)$, or equivalently the ratio 
$R=m_\nu / m_N \sim {\cal O}(10^{-11})$ as long as the ratios of 
masses of the particles propagating in the loop satisfy certain relations.
In Ref. \cite{PQ} it has been shown that, taking the masses in units of 
$M_F$, where $M_F$ could be of ${\cal O}(\geq 200GeV)$,  one obtains 
$R \lesssim 10^{-11}$ for 
$M_G/M_{{\cal M}_2} \lesssim 10^{-3}$ when $M_G \lesssim 10^5$, 
or for $M_G/M_{{\cal M}_2} \sim 10^{-2} - 10^{-1}$ when $M_G > 10^7$, with 
$M_P \sim 1 - 10^2$.
After lifting the degeneracy of the three light neutrinos,  
by breaking the remaining family symmetry $SO(3)$, one obtains the following 
mass eigenvalues for the three light neutrinos

\begin{equation}
m_1 = m_N\frac{\sin(2\beta)}{32\,\pi^2}\{\Delta I(G,P)-
b\Delta I(G,P,b)\} \, ,
\end{equation}

\begin{equation}
m_2 = m_N \frac{\sin(2\beta)}{32\,\pi^2}\Delta I(G,P) \, , 
\end{equation}

\begin{equation}
m_3 = m_N \frac{\sin(2\beta)}{32\,\pi^2}\{\Delta I(G,P)+
b\Delta I(G,P,-b)\} \, ,
\end{equation}  

\noindent
where $\Delta I(G,P,\pm b)$ is given by

\begin{equation}
\Delta I(G,P,\pm b) \equiv I(M_G, \pm b)- I(M_P, \pm b) \, ,
\label{deltaI}
\end{equation}

\noindent
with 

\begin{eqnarray}
I(M_G, \pm b) & = & \frac{M_G^2}{M_F-M_{{\cal M}_2}}
\{\frac{M_F[-M_F^2 (1 \pm b + \ln(\frac{M_G^2}{M_F^2})) + M_G^2 (1 \pm b)]}
{(M_G^2-M_F^2)^{2}(1 \pm b(M_G^2/(M_G^2-M_F^2)))} -(M_F 
\leftrightarrow M_{{\cal M}_2})\}\, . \nonumber \\ 
\end{eqnarray}

The mass splittings, neglecting terms of order $b^2$ are now 

\begin{equation}
m_3^2 - m_2^2 = m_2  m_N 2 b \frac{\sin(2\beta)}{32\,\pi^2} 
\Delta I(G,P,-b) \, ,  
\end{equation}
\begin{equation}
m_2^2 - m_1^2 = m_2  m_N 2 b \frac{\sin(2\beta)}{32\,\pi^2}
 \Delta I(G,P,b)  \, .
\end{equation} 

\noindent
The above mass splittings are almost degenerate and in Ref. \cite{PQ}  
a possible solution to lift this degeneracy was presented. 
What has been showed is that 
if one introduces mixing terms in the neutrino mass matrix between the 
three light families and the $4^{th}$ family, one can obtain reasonable values 
for the mass splittings. 
Two numerical examples have been presented. The mass splittings 
for the 
particular choice of the parameter $b=0.000095$ that satisfies the upper 
constraint for $b$ derived in this paper and presented in Eq. \ref{bconstr} 
are given in Eq. \ref{masssplit1} and \ref{masssplit2}.

\section{Appendix B}

In the following we will show in a simple example how the dependence on 
the parameter $b$, introduced in Eq. \ref{massomega}, for the amplitude of 
the FCNC processes in our model appears. This example will be introduced 
in an analagous fashion to a FCNC process in the SM, which we know to be 
governed by the GIM mechanism. 
The process that we are considering is $ t \rightarrow c W^+ W^-$. 
Now the amplitude of this process is proportional to the quantity

\begin{equation}
\sum_{j=d,s,b}V_{tj}^*V_{cj}\frac{1}{p^2-m_j^2} \, ,
\label{ampSM}
\end{equation}

\noindent
where the sum appears because one has to consider all possible virtual 
down-quarks, and $p$ is the momentum of the virtual quarks. Notice that
Eq. \ref{ampSM} represents the approximation $p \gg m_j$ which
is a valid one as we shall see below.
$V_{ij}$ is the CKM matrix and the unitarity of this matrix implies in 
particular the relation

\begin{equation}
\sum_{j=d,s,b}V_{tj}^*V_{cj} = 0 \, .
\end{equation}

\noindent
Using the above relation we can write Eq. \ref{ampSM} as

\begin{eqnarray}
 -\sum_{j=s,b}V_{tj}^*V_{cj}\frac{1}{p^2-m_d^2} + 
\sum_{j=d,s,b}V_{tj}^*V_{cj}\frac{1}{p^2-m_j^2} & = &  
\sum_{j=d,s,b}V_{tj}^*V_{cj}\frac{m_j^2-m_d^2}{(p^2-m_d^2)(p^2-m_j^2)} \, .
\nonumber \\  
\end{eqnarray}

\noindent
Using now the approximation

\begin{equation}
V_{ts}^*V_{cs} \simeq - V_{tb}^*V_{cb} \approx -{\theta}^2 \, ,
\end{equation}

\noindent
and doing the sum we can rewrite Eq. \ref{ampSM} as 

\begin{equation}
{\theta}^2\frac{m_b^2-m_s^2}{(p^2-m_s^2)(p^2-m_b^2)} \, .
\end{equation}

\noindent
From kinematic considerations $p^2$ falls in the range 

\begin{equation}
(M_W+m_c)^2 < p^2 < (m_t-M_W)^2 \, ,
\end{equation}

\noindent
which means that $p^2$ is of order of $M_W^2$.
One obtains the final expression

\begin{equation}
{\theta}^2\frac{m_b^2}{M_W^2}\frac{1}{p^2-m_b^2} \, ,
\end{equation}

\noindent
which shows the characteristic term of the GIM mechanism $m_b^2/M_W^2$ for 
FCNC processes involving external $t$. 

Now, looking at our model, and assume that the process 
$t \rightarrow c F {\cal M}_1$ 
can kinematically occur, (although it cannot be in reality because 
$M_F \geq 200 GeV$ and $M_{{\cal M}_1} \geq M_F$).
The amplitude will be proportional to 

\begin{equation}
\sum_{j}A_{\Omega,j3}A_{\Omega,2j}^T\frac{1}{k^2-M_{\Omega_j}^2} \, ,
\label{ampPQ}
\end{equation}

\noindent
where $A_{\Omega}$ is a unitary matrix given by (see Ref. \cite{PQ})

\begin{equation}
A_{\Omega} = \left(\begin{array}{cccc}
\frac{1}{2}  & \frac{1}{\sqrt{2}} & \frac{1}{2} \\
\frac{1}{2} & -\frac{1}{\sqrt{2}} & \frac{1}{2} \\
 -\frac{1}{\sqrt{2}} &  0  & \frac{1}{\sqrt{2}}  
\end{array}\right) \, ,
\label{matrixAnew} 
\end{equation}

\noindent
chosen different from that one given in Eq. \ref{matrixA} because 
otherwise we will have no transition from top to charm, 
and the sum comes out from the fact that we have to consider as virtual states 
any gauge boson of $SO(3)$.
The unitarity of the matrix $A_{\Omega}$ implies the particular relation 

\begin{equation}
\sum_{j}A_{\Omega,j3}A_{\Omega,2j}^T = 0 \, ,
\end{equation}

\noindent 
that we can use to rewrite the amplitude as 

\begin{equation}
-\sum_{j=2,3}A_{\Omega,j3}A_{\Omega,2j}^T\frac{1}{k^2-M_{\Omega_1}^2} +
\sum_{j=2,3}A_{\Omega,j3}A_{\Omega,2j}^T\frac{1}{k^2-M_{\Omega_j}^2} \, ,
\end{equation}

\noindent
in the same way as we did for the SM. 
Now for the particular matrix $A_{\Omega}$ that we have chosen, we obtain

\begin{eqnarray}
A_{\Omega,23}A_{\Omega,22}^T & = & -\frac{1}{2\sqrt{2}} \, , \nonumber \\
A_{\Omega,33}A_{\Omega,23}^T & = & 0 \, ,
\end{eqnarray}

\noindent
which gives

\begin{equation}
-\frac{1}{2\sqrt2}\frac{M_{\Omega_2}^2-M_{\Omega_1}^2}{(k^2-M_{\Omega_1}^2)
(k^2-M_{\Omega_2}^2)} \, .
\end{equation}

\noindent
Substituting for $M_{\Omega_1}^2$ and $M_{\Omega_2}^2$ the expressions given 
in Eq. \ref{massomega} we obtain

\begin{equation}
\frac{1}{\sqrt2}\frac{b\,M_G^2}{(k^2-M_G^2(1+b))
(k^2-M_G^2(1-b))}\, .
\end{equation}
 
\noindent
From kinematically considerations we have the following constraints for $k^2$

\begin{equation}
(m_c+M_{M_1})^2 < k^2 < (m_t-M_F)^2 \, ,
\end{equation}

\noindent
which implies $k^2 \ll M_G^2$, with $ M_{{\cal M}_1}^2, M_F^2 \ll M_G^2 $. 
The final expression becomes

\begin{equation}
\frac{1}{\sqrt2}\frac{b}{k^2-M_G^2} \, ,
\end{equation}

\noindent
which shows the linear dependence on the parameter $b$.
One notices that the linear dependance on the parameter $b$ 
comes out in the same way as the linear dependence on the quantity 
$m_b^2/M_W^2$ in the GIM mechanism of the SM. 
It is important to notice here that, in the SM, the GIM suppression
comes from two factors: the appropriate product of CKM elements
and the ratio $m_{q}^2/m_{W}^2$, where $m_q$ is a quark mass (up or down).
(Even if $m_{q}^2/m_{W}^2 \sim {\cal O}(1)$ such as for the top quark,
the CKM suppression factor can be very small.)  
In our case, the amplitude is always suppressed 
by the parameter $b$.

\begin{figure}
\caption{Feynman graphs for the effective verteces of the $Z$ and $\gamma$ for 
$\Delta S=1$ processes $s \rightarrow d$. 
The self energy diagrams are omitted.}
\end{figure}

\begin{figure}
\caption{Feynman graphs for the effective vertex of the $g$ for 
$\Delta S=1$ processes $s \rightarrow d$. The self energy 
diagrams are omitted.}
\end{figure}

\begin{figure}
\caption{The function $C_0(x_F,b)$, appearing in the amplitude for the $Z$ 
effective vertex is plotted in linear and logarithmic scale for $b=0.001$ 
and $ b=0.0001 $. The function $C_0(x_F)$,  which is the first 
non vanishing coefficient in the Taylor expansion in $b$ of 
$C_0(x_F,b)$  is also plotted.}
\end{figure}

\begin{figure}
\caption{The logarithm of the function $D_0(x_F,b)$, appearing in the 
amplitude for the $\gamma$ 
effective vertex is plotted versus the logarithm of $x_F$ for $b=0.001$ 
and $ b=0.0001 $. The function $D_0(x_F)$, which is the first non vanishing 
coefficient in the Taylor expansion in $b$ of 
$D_0(x_F,b)$ is also plotted.}
\end{figure}
 
\begin{figure}
\caption{Box diagrams for $\Delta S=1$ process $sd \rightarrow dd$. 
The corresponding box diagrams with the pseudo NG bosons 
$Re \tilde{\rho}_i^{\prime}$ instead of the NG bosons 
$\tilde{\Omega}_i^{\prime}$ are omitted.}
\end{figure}
 
\begin{figure}
\caption{The function $B_0(x_F,b)$, appearing in the amplitude for the box
diagrams in the process $sd \rightarrow dd$ is plotted in linear and 
logarithmic scale for $b=0.001$ 
and $ b=0.0001 $. The function $B_0(x_F)$, which is the first non vanishing 
coefficient in the Taylor expansion in $b$ of 
$B_0(x_F,b)$ is also plotted.}
\end{figure}

\begin{figure}
\caption{The function $B_{mix0}(x_F,x_M,b)$, appearing in the amplitude 
for the box
diagrams in the process $sd \rightarrow dd$ is plotted in linear and 
logarithmic scale for $b=0.001$ 
and $ b=0.0001 $ versus $x_M$, with the condition $x_F=10^{-4}x_M$.}
\end{figure}

\begin{figure}
\caption{Feynman graphs for the 
$\Delta S=1$ process $sd \rightarrow \mu e$. In the graph $(e)$ the  
vertices are effective. The corresponding diagrams with the pseudo NG bosons
$Re \tilde{\rho}_i^{\prime}$ instead of the NG bosons 
$\tilde{\Omega}_i^{\prime}$ are omitted.}
\end{figure}

\begin{figure}
\caption{The logarithm of the function $(C_{Lq}-C_{Rq})^2(C_{Ll}^2+C_{Rl}^2)$, 
which appears in Eq. \protect\ref{BRKLmue} for the $BR(K_L \rightarrow \mu e)$ 
is plotted versus  
$Log(x_M/x_F)$, for $b=2.5\cdot10^{-4}$, $M_G=200TeV$ and $x_F=10^{-6}$.}
\end{figure}

\begin{figure}
\caption{The logarithm of $BR(K_L \rightarrow \mu e)$ 
is plotted versus  
$Log(x_M/x_F)$, for $b=2.5\cdot10^{-4}$, $M_G=200TeV$ and $x_F=10^{-6}$.
The lowest flat zone corresponds to $BR(K_L \rightarrow \mu e)=
6.65\cdot10^{-22}$, while the highest flat zone corresponds to 
$BR(K_L \rightarrow \mu e)=1.02\cdot10^{-14}$.}
\end{figure}

\begin{table}
\caption{Particle content and quantum numbers of
$SU(3)_c \otimes SU(2)_L \otimes U(1)_Y \otimes SO(N_f) \otimes SU(2)_{\nu_R}$}
\begin{tabular}{l|r}
Standard Fermions      & $q_L = (3, 2, 1/6, N_f, 1)$\\
                       & $l_L = (1, 2, -1/2, N_f, 1)$\\
                       & $u_R = (3, 1, 2/3, N_f, 1)$\\
                       & $d_R = (3, 1, -1/3, N_f, 1)$\\
                       & $e_R = (1, 1, -1, N_f, 1)$\\ \hline
Right-handed $\nu$'s   & Option 1: $\eta_R = (1, 1, 0, N_f, 2)$ \\
                       & Option 2: $\eta_R = (1, 1, 0, N_f, 2)$;  \\ 
                       & $\eta_R^{\prime} = (1, 1, 0, 1, 2)$ \\ \hline
Vector-like Fermions   & $F^l_{L,R} = (1, 2, -1/2, 1, 1)$\\ 
for the lepton sector  & ${\cal M}^l_{1 L,R} = (1, 1, -1, 1, 1)$\\
                       & ${\cal M}^l_{2 L,R} = (1, 1, 0, 1, 1)$\\ \hline
Vector-like Fermions   & $F^q_{L,R} = (3, 2, 1/6, 1, 1)$\\ 
for the quark sector   & ${\cal M}^q_{1 L,R} = (3, 1, -1/3, 1, 1)$\\
                       & ${\cal M}^q_{2 L,R} = (3, 1, 2/3, 1, 1)$\\ \hline
Scalars                & $\Omega^{\alpha} = (1, 1, 0, N_f, 1)$\\
                       & $\rho_{i}^{\alpha} = (1, 1, 0, N_f, 2)$ \\
                       & $\phi = (1, 2, 1/2, 1, 1)$ 
\end{tabular}
\end{table}

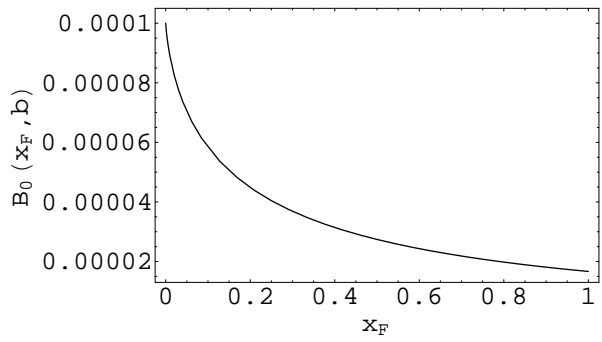
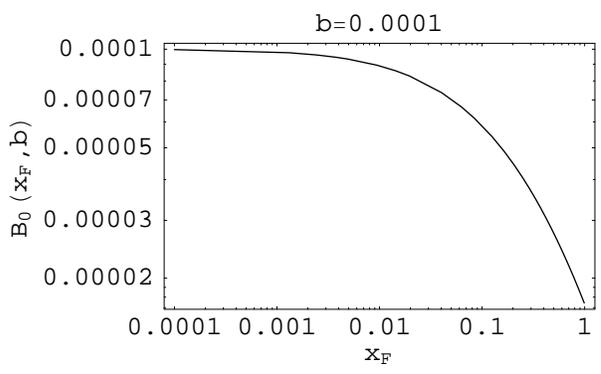
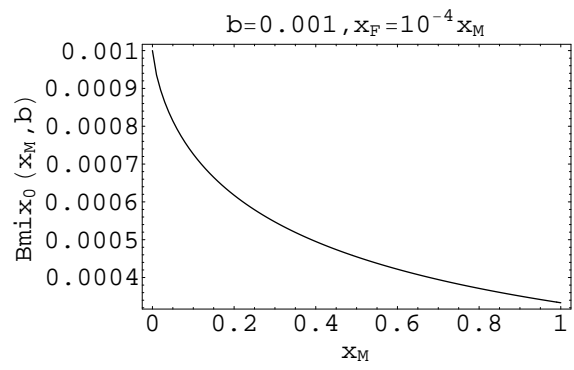
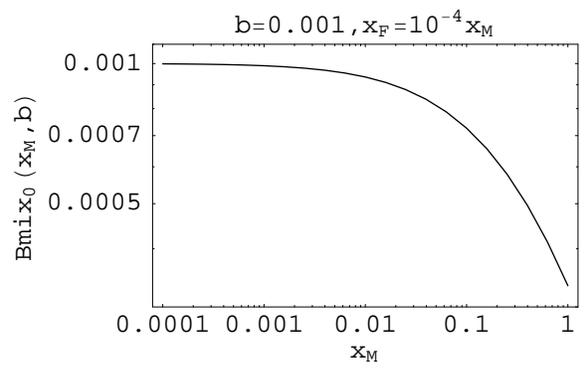
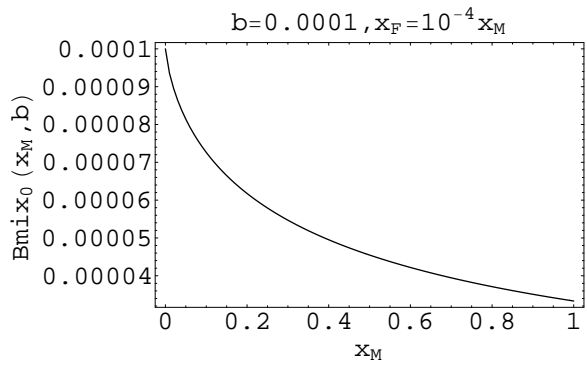
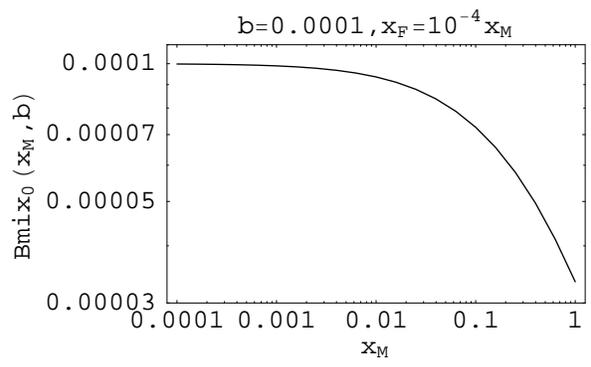
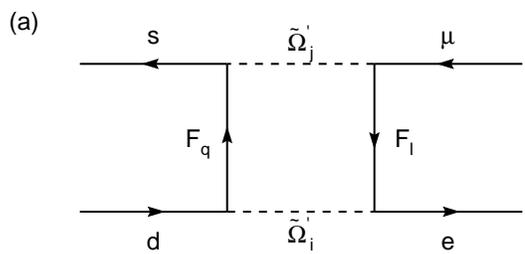
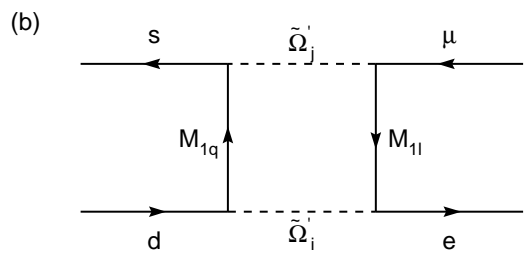
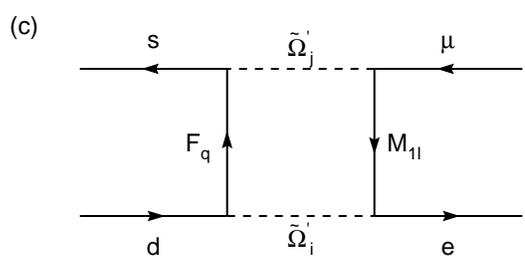
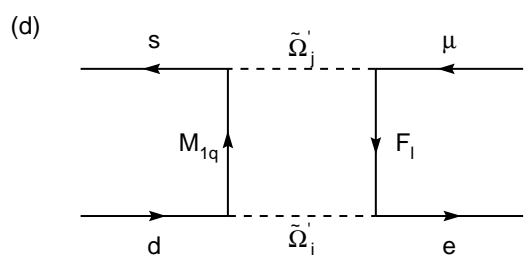
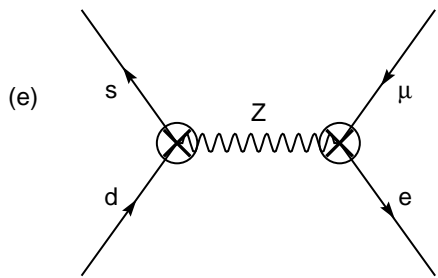
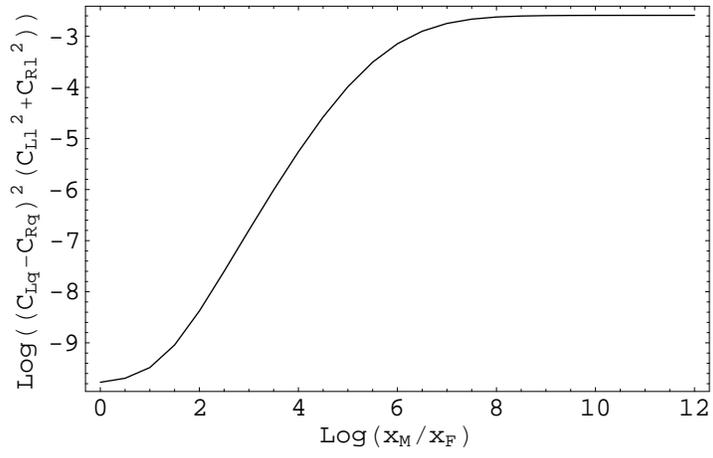
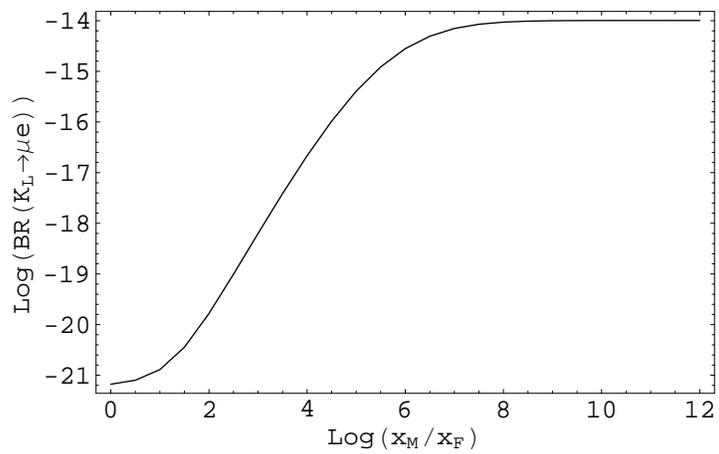
\end{document}